\documentclass[]{spie}  

\usepackage{amsmath,amsfonts,amssymb}
\usepackage{graphicx}
\usepackage[colorlinks=true, allcolors=blue]{hyperref}
\usepackage{float}
\usepackage{subcaption}
\usepackage{hyperref}
\usepackage{booktabs}

\renewcommand{\vec}[1]{\mathbf{#1}}

\usepackage{setspace}

\title{Advancing the AmbientGAN for learning \\stochastic object models}

\author[a]{Weimin Zhou}
\author[b]{Sayantan Bhadra}
\author[c]{Frank J. Brooks}
\author[d]{Jason L. Granstedt}
\author[c, e]{\\Hua Li}
\author[c]{Mark A. Anastasio}
\affil[a]{Department of Psychological and Brain Sciences,
\break University of California, Santa Barbara, Santa Barbara, CA 93106, USA}
\affil[b]{Department of Computer Science and Engineering,
\break Washington University in St$.\ $Louis, St$.\ $Louis, MO 63130, USA}
\affil[c]{Department of Bioengineering, 
\break University of Illinois at Urbana-Champaign, Urbana, IL 61801, USA}
\affil[d]{Department of Computer Science, 
\break University of Illinois at Urbana-Champaign, Urbana, IL 61801, USA}
\affil[e]{Carle Cancer Center,
\break Carle Foundation Hospital, Urbana, IL 61801, USA}

\authorinfo{Further author information: (Send correspondence to Mark A. Anastasio)\\Mark A. Anastasio: E-mail: maa@illinois.edu, Telephone: 1 217 333 1867}

\begin{document} 
\maketitle

\begin{abstract}
Medical imaging systems are commonly assessed and optimized by use of objective-measures of image quality (IQ)
that quantify the performance of an observer at specific tasks. 
Variation in the objects to-be-imaged is an important source of variability that can
significantly limit observer performance.
This object variability can be described by
stochastic object models (SOMs).
In order to establish SOMs that can accurately model realistic object variability, 
it is desirable to use experimental data.
To achieve this, an augmented generative adversarial network (GAN) architecture called AmbientGAN
has been developed and investigated.
However, AmbientGANs cannot be immediately trained by use of advanced GAN training methods such as the progressive growing of GANs (ProGANs). Therefore, the ability of AmbientGANs to establish realistic object models is limited.
To circumvent this,
a progressively-growing AmbientGAN (ProAmGAN)
has been proposed.
However, 
ProAmGANs are designed for generating two-dimensional (2D) images while
medical imaging modalities are commonly employed for imaging
three-dimensional (3D) objects.
Moreover, ProAmGANs that employ traditional generator architectures
lack the ability to control specific image features such as fine-scale textures that are frequently considered when optimizing imaging systems.
In this study, we address these limitations by
proposing two advanced AmbientGAN architectures:
3D ProAmGANs
and Style-AmbientGANs (StyAmGANs).
Stylized numerical studies involving magnetic resonance (MR) imaging systems are conducted.
The ability of 3D ProAmGANs to learn 3D SOMs from imaging measurements
and the ability of StyAmGANs to control fine-scale textures of synthesized objects are demonstrated.
\end{abstract}

\keywords{stochastic object model, generative adversarial networks, signal detection, objective assessment of image quality}

\section{INTRODUCTION}
\label{sec:intro}  
It has been advocated that medical imaging systems should be assessed and optimized by use of objective measures of image quality (IQ)
that quantify the performance of an observer at specific tasks\cite{barrett2013foundations, zhou2019approximating, zhou2020approximating, zhou2020markov}.
To achieve this, all sources of randomness in the imaging measurements should be accounted for\cite{barrett2013foundations}.
One important source of randomness that can limit observer performance is the variation in the ensemble of objects to-be-imaged\cite{barrett2013foundations,kupinski2003experimental}.
To describe this randomness, 
a stochastic object model (SOM) can be established that can generate an ensemble of objects that have prescribed statistical properties.
In order to establish a SOM that can capture realistic variations of object textures and anatomical structures,
it is desirable to use experimental data. 

Generative adversarial networks (GANs) hold great potential to learn SOMs.
However, traditional GANs that are typically trained by use of reconstructed images are influenced by the effects of the reconstruction process and the measurement noise.
To circumvent this, we investigated an augmented GAN architecture called an AmbientGAN \cite{bora2018ambientgan}
to establish a simple lumpy object model\cite{zhou2019learning}. 
However, AmbientGANs cannot be readily implemented with advanced training procedures such as the progressive growing of GAN~(ProGAN)~\cite{karras2017progressive}. Therefore, the ability
of conventional AmbientGANs to establish realistic SOMs is limited.

Recently, we proposed an AmbientGAN training method named progressively-growing AmbientGAN (ProAmGAN)\cite{zhou2020progressively, zhou2020learning}
by augmenting the original ProGAN with a measurement operator and a reconstruction operator.
However, ProAmGANs are designed for learning two-dimensional (2D) SOMs while medical imaging systems are typically 
employed for imaging three-dimensional (3D) objects.
Therefore,
 there is still a need to develop a method for establishing realistic 3D SOMs.
 
Image textures are often considered in the design and optimization of imaging systems.
For example, lumpy background and clustered lumpy background models\cite{barrett2013foundations} have been proposed.
When fine-scale features are considered in the design and optimization of imaging systems,
it may be useful to generate objects that have the same large-scale structures but have different fine-scale features.
Additionally,
the ability to control scale-specific image features
can potentially benefit generative model-based reconstruction methods\cite{bhadra2020medical, kelkar2020compressible} for dynamic imaging
because it enables image reconstruction
under constraints on scale-specific image structures.
However, ProAmGANs cannot achieve this because they 
employ traditional generator architectures that lack the ability to control features of synthesized images.  
 
In this study, we address these limitations by proposing two advanced AmbientGAN architectures: (1) a 3D ProAmGAN that employs a 3D ProGAN architecture\cite{eklund2019feeding} for learning 3D SOMs from imaging measurements and (2) a novel Style-AmbientGAN (StyAmGAN) that employs a style-based generator \cite{karras2019style, karras2019analyzing} for controlling styles and features of
the AmbientGAN-synthesized images.
Stylized numerical studies involving magnetic resonance (MR) imaging systems are conducted.
It is demonstrated that the 3D ProAmGAN can successfully establish SOMs for generating images that contain $128\times 128 \times 128$ voxels 
and the StyAmGAN can provide the ability to control specific image features.
{Combining the two proposed AmbientGAN architectures, a 3D StyAmGAN can also be developed to establish controllable 3D SOMs from imaging measurements.}

\section{Background}
This study considers linear imaging systems
that can be described as: $\vec{g} = H\vec{f}+\vec{n}$.
Here, $\vec{g}\in \mathbb{R}^{N}$ denotes the measured image data,
$\vec{f}\in \mathbb{R}^{M}$ denotes a finite-dimensional representation of objects to-be-imaged,
$H\in \mathbb{R}^{N\times M}$ denotes a discrete-to-discrete (D-D) imaging operator, and $\vec{n} \in \mathbb{R}^{N}$ denotes the measurement noise.
Below, previous work on AmbientGANs  is reviewed and the 3D ProAmGAN and Style-AmbientGAN are developed.

\subsection{AmbientGANs}
An AmbientGAN \cite{bora2018ambientgan} comprises a generator and a discriminator that are both represented by a deep neural network.
The generator maps a latent vector $\vec{z}\in \mathbb{R}^k$ to a generated image $\hat{\vec{f}}\in \mathbb{R}^M$: $\hat{\vec{f}} = G(\vec{z}; \Theta_G)$, where $G(\cdot; \Theta_G)$ denotes the
mapping function of the generator that are parameterized by a set of weight parameters $\Theta_G$.
The measurement operator $\mathcal{H}_\vec{n}$ is subsequently employed
to simulate the imaging measurement data $\hat{\vec{g}}$: $\hat{\vec{g}} = \mathcal{H}_\vec{n} (\hat{\vec{f}}) \equiv H\hat{\vec{f}}+\vec{n}$.
The discriminator that
is represented by another deep neural network 
having a mapping function $D: \mathbb{R}^N \rightarrow \mathbb{R}$,
which is parameterized by a set of parameters $\Theta_D$,
maps the experimental imaging measurements $\vec{g}$ and AmbientGAN-simulated imaging measurements $\hat{\vec{g}}$
to a real-valued scalar $s$. This value is employed to distinguish the experimental and simulated imaging measurements.
The AmbientGAN training can be represented by 
a two-player minimax game:
\begin{equation} \label{eq:GAN}
\min_{\Theta_G} \max_{\Theta_D} {E_{\vec{g} \sim p_g}} [l\left(D(\vec{g}; \Theta_D)\right)] + {E_{\vec{z}\sim p_z}} [l(1- D\left(\hat{\vec{g}}; \Theta_D) \right)],
\end{equation}
where $\hat{\vec{g}} = \mathcal{H}_\vec{n} ( G(\vec{z}; \Theta_G))$ and $l(\cdot)$ represents a loss function employed in the training process.
Bora \emph {et al.} have shown that when the probability density function (pdf) $p_\vec{f}$ uniquely induces the pdf $p_\vec{g}$, 
the pdf of the AmbientGAN-generated images $p_{\hat{\vec{f}}}$ is identical to the ground-truth $p_\vec{f}$ when 
the global optimum of the two-player minimax game is achieved \cite{bora2018ambientgan}.

\subsection{Progressively Growing AmbientGANs}
In order to stably train AmbientGANs for establishing more realistic and complicated object models,
more recently,
a progressively-growing AmbientGAN (ProAmGAN) was developed\cite{zhou2020learning, zhou2020progressively}.
The ProAmGAN architecture augments the original Progressively-growing GAN (ProGAN)
with the measurement operator $\mathcal{H}_\vec{n}: \mathbb{R}^{M}\rightarrow \mathbb{R}^N$ and a reconstruction operator $\mathcal{O}:  \mathbb{R}^{N}\rightarrow \mathbb{R}^M$.
In the ProAmGAN training process,
the generator is trained to produce objects $\hat{\vec{f}}$
such that the corresponding reconstructed images $\hat{\vec{f}}_{Recon} \equiv \mathcal{O}( \mathcal{H}_\vec{n}(\hat{\vec{f}}) )$ 
can mimic the reconstructed images $\vec{f}_{Recon} = \mathcal{O}(\vec{g})$ corresponding to the experimental imaging measurements $\vec{g}$.
Numerical studies considering a variety of stylized medical imaging systems in combination with different 
object ensembles were conducted previously to demonstrate the ability of the ProAmGAN to learn 2D SOMs\cite{zhou2020learning}.

\section{3D Pro-AmbientGANs and Style-AmbientGANs} 
\begin{figure}[H]
\centering
{\includegraphics[width=\linewidth]{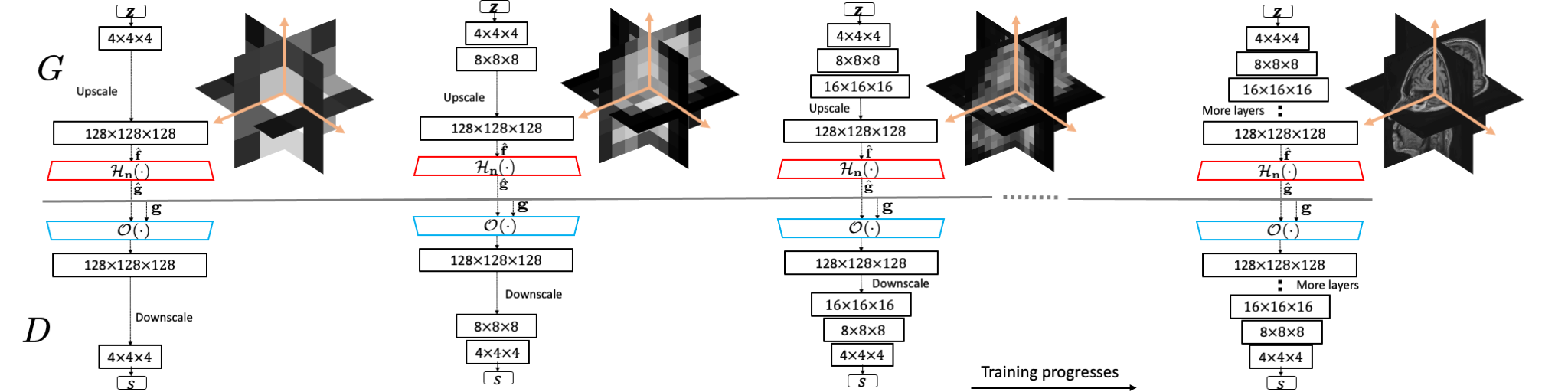}}
\vspace{-.2cm}
\caption{3D ProAmGAN training process. The training starts with the resolution of $4\times 4\times 4$. The resolution is doubled gradually by adding more layers to the generator and the discriminator until the final resolution is achieved.}
\label{fig:arc}
\end{figure}

Recently, a 3D ProGAN was developed by Eklund based on the original 2D ProGAN\cite{eklund2019feeding}. 
In this study, we implement this 3D ProGAN and augment it to the 3D ProAmGAN by use of the same strategy employed in the 2D ProAmGAN.
The training process of the 3D ProAmGAN is illustrated in Fig. \ref{fig:arc}.

Karras \emph{et al.} recently developed a StyleGAN architecture \cite{karras2019style, karras2019analyzing}
that can disentangle the latent factors of the object variation and subsequently enables the control of specific image styles and features of the synthesized images.
Accordingly, it may provide a way to
generate objects that have the same large-scale structures but different fine-scale features. 
Details on the StyleGAN architecture and its improved version, i.e., StyleGAN2, can be found in the literature\cite{karras2019style, karras2019analyzing}.
In this study, we propose a Style-AmbientGAN (StyAmGAN) by augmenting 
the StyleGAN or StyleGAN2 architecture with the measurement operator $\mathcal{H}_\vec{n}$ and the reconstruction operator $\mathcal{O}$.
As with the ProAmGAN, the generator in the StyAmGAN learns to generate images $\hat{\vec{f}}$ such that the corresponding 
reconstructed images $\hat{\vec{f}}_{Recon} \equiv \mathcal{O}( \mathcal{H}_\vec{n}(\hat{\vec{f}}) )$ can mimic the reconstructed images ${\vec{f}}_{Recon}\equiv \mathcal{O}(\vec{g})$ corresponding to the training measurement data $\vec{g}$. After training the StyAmGAN, the style-based generator can be employed to produce images and control specific image styles and features.

\section{Numerical studies}
\subsection{3D ProAmGAN} 
A stylized 3D MR imaging system that fully samples k-space data was considered. 
A 3D brain dataset from
Alzheimer's Disease Neuroimaging Initiative (ADNI) (\url{http://adni.loni.usc.edu/data-samples/})
was employed to serve as a set of ground-truth objects.
 Six hundred 3D volumes were selected from this dataset and resized to the dimension of $128\times 128 \times 128$.
 A set of 600 measurement data was simulated by adding complex Gaussian noise to 
the 3D discrete Fourier transform (DFT) of the object $\vec{f}$.
The reconstruction operator $\mathcal{O}(\cdot)$
was a 3D inverse DFT to be used for computing the reconstructed image $\vec{f}_{Recon}$ and $\hat{\vec{f}}_{Recon}$.
The 3D ProGAN code (\url{https://github.com/wanderine/ProgressiveGAN3D}) was implemented and modified according to the architecture illustrated in Fig. \ref{fig:arc}.
The training started with a resolution of $4\times 4 \times 4$, and the resolution along each dimension was 
doubled during the training process until the final resolution of $128\times 128\times 128$ was achieved.
The training of the 3D ProAmGAN took about 15 days by use of 4 Nvidia Quadro RTX 8000 GPUs.

The Fréchet Inception Distance (FID) score was employed to evaluate the performance of the 3D ProAmGAN. 
Smaller FID score indicates better quality of the generative model.
Three different FID scores were computed  by use of sagittal slices, coronal slices and axial slices, respectively. 
In addition, a task-based validation study considering a signal detection task was conducted. 
The imaging processes under the signal-absent hypothesis ($H_0$) and the signal-present hypothesis ($H_1$) can be described as:
\begin{subequations}
\label{eq:imgH_s}
\begin{align}
H_{0}:&\  \mathbf{g} = \mathbf{f} + \mathbf{n}, \\
H_{1}:&\  \mathbf{g} = \mathbf{f} + \mathbf{s} + \mathbf{n},
\end{align}
\end{subequations}
where $\vec{f}$ denotes the object to-be-imaged, $\vec{s}$ is a 3D sphere signal having the radius of 2 voxels, and $\vec{n}$ is independent and identically distributed (i.i.d.) Gaussian noise.
The signal detection performance 
was quantified by the signal-to-noise ratio of the Hotelling observer ($SNR_{HO}$) \cite{barrett2013foundations}:
\begin{equation}
SNR_{HO} = \sqrt{\vec{s}^{T}\mathbf{K}^{-1}\vec{s}},
\end{equation}
where $\mathbf{K}$ denotes the covariance matrix of $\vec{g}$. The $SNR_{HO}$ was evaluated on a region of interest (ROI) of dimension $8\times 8\times 8$
centered at the signal.

As a comparison, the original 3D ProGAN\cite{eklund2019feeding} was trained by use of the reconstructed volumes $\vec{f}_{Recon}$. The generator in the 3D ProGAN
was trained to learn the distribution of the noisy reconstructed volume $\vec{f}_{Recon}$ instead of establishing the SOM that describes the distribution of the object $\vec{f}$.

\subsection{Style-AmbientGANs} 
A stylized 2D MR imaging system
was considered.
A collection of 30,000 sagittal slices 
was extracted from ADNI dataset and resized to the dimension of $128\times 128$ to serve as the ground-truth objects.
 The imaging measurements were simulated by computing the 2D DFT 
 and adding a complex Gaussian noise.
 The reconstruction operator $\mathcal{O}(\cdot)$
was a 2D inverse DFT.
 The Ambient-StyleGAN was implemented by modifying the code of StyleGAN2 (\url{https://github.com/NVlabs/stylegan2}) by adding the corresponding measurement operator
 $\mathcal{H}_\vec{n}(\cdot)$ and reconstruction operator $\mathcal{O}(\cdot)$ to the original StyleGAN2 architecture.

 \section{Results}
\subsection{3D ProAmGAN} 
\begin{figure}[H]
\centering
{\includegraphics[width=\linewidth]{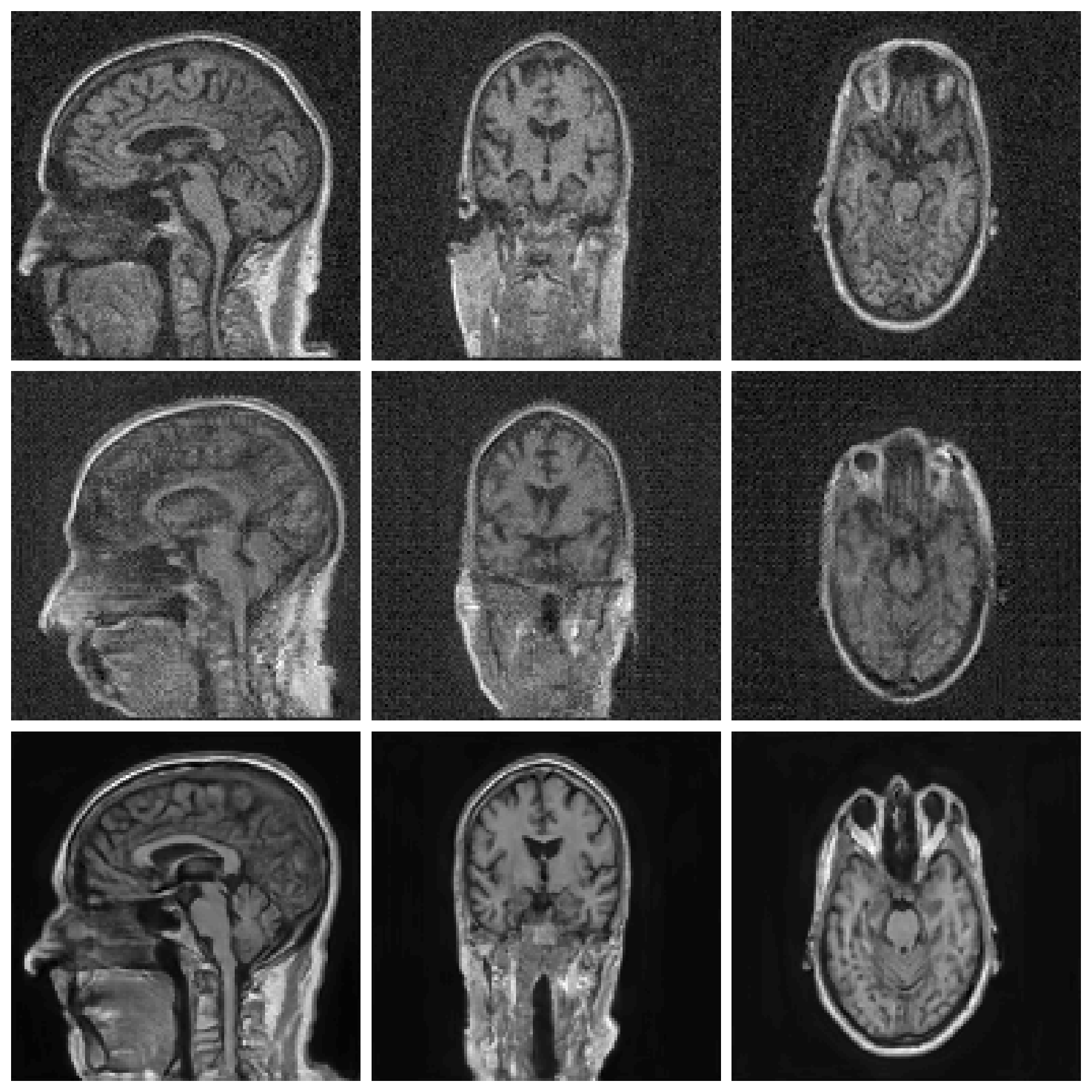}}
\caption{Top row: a noisy reconstructed volume $\vec{f}_{Recon}$ of the training data $\vec{g}$; Middle row: a ProGAN-generated volume; Bottom row: a ProAmGAN-generated volume. 
Images from left to right correspond to a sagittal slice, a coronal slice, and an axial slice of the 3D volume, respectively. }
\label{fig:real_fake}
\end{figure}

Sagittal, coronal, and axial slices of a reconstructed volume $\vec{f}_{Recon}$ corresponding to the training measurement data $\vec{g}$~
are shown in the top row of Fig. \ref{fig:real_fake}.
Examples of sagittal, coronal, and axial slices produced by the 3D ProGAN and 3D ProAmGAN are shown in the middle row and bottom row of Fig. \ref{fig:real_fake}, respectively.
The volumes produced by the ProGAN that was trained on the noisy reconstructed volumes $\vec{f}_{Recon}$
are significantly affected by the measurement noise;
while the ProAmGAN-generated volumes $\hat{\vec{f}}$
are clean. This is because the ProAmGAN was trained to learn the SOM from the measurement data and a known measurement model;
while the ProGAN was trained to learn the distribution of the noisy reconstructed volumes.
The results demonstrate the ability of ProAmGANs to mitigate the measurement noise when establishing 3D SOMs.

\begin{figure}[H]
\centering
\includegraphics[width=\linewidth]{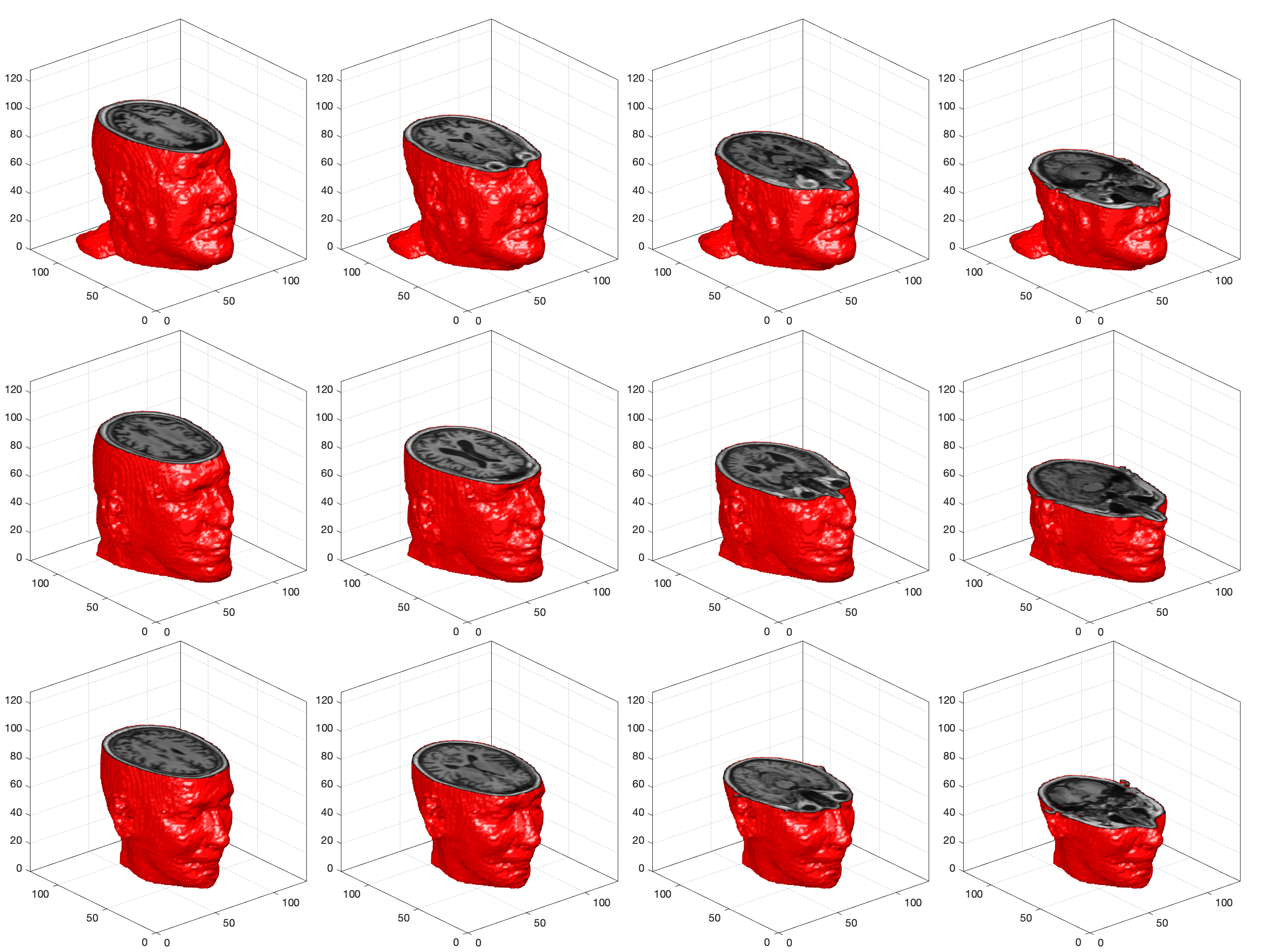}
\caption{3D visualization of three ProAmGAN-generated 3D volumes. Each row corresponds to a ProAmGAN-generated 3D volume.}
\label{fig:fake_3D}
\end{figure}

Three-dimensional visualizations of three ProAmGAN-generated volumes are shown in Fig. \ref{fig:fake_3D}. In this figure, each row plots axial slices from a ProAmGAN-generated 3D volume with the corresponding isosurfaces that depict the shape of the 3D volume. 
The ProAmGAN-generated 3D volumes are visually promising.
The FID scores evaluated on axial, coronal, and sagittal slices are summarized in Table. \ref{table:metric}.
The ProAmGAN-generated volumes produced smaller FID scores than those produced by the ProGAN.
This indicates that the ProAmGAN outperformed the ProGAN in terms of generating high-quality 3D volumes.
The signal detection performance that is quantified by $SNR_{HO}$ is also shown in Table. \ref{table:metric}.
It is observed that the ProAmGAN
can estimate the ground truth $SNR_{HO}$  more accurately than the ProGAN for the considered signal detection task.
\begin{table}[H]
\center
\begin{tabular}{l|c|c|c|c}
\hline\hline
              & \multicolumn{1}{c|}{\begin{tabular}[c]{@{}c@{}}FID:\\ Axial\end{tabular}} & \multicolumn{1}{c|}{\begin{tabular}[c]{@{}c@{}}FID:\\ Coronal\end{tabular}} & \begin{tabular}[c]{@{}c@{}}FID:\\ Sagittal\end{tabular} & \begin{tabular}[c]{@{}c@{}}$SNR_{HO}$:\\ Reference 1.7039\end{tabular} \\ \hline
3D ProGAN                     & 86.9011                                                                   & 66.0846                                                                     & 77.6631                                                 & 1.4720                              \\ \hline
\textbf{3D ProAmGAN}   &  \textbf{32.7899}                                                          & \textbf{32.0720}                                                            & \multicolumn{1}{l|}{\textbf{39.8564}}                   & \textbf{1.8101}                              \\ \hline
\end{tabular}
\vspace{0.1cm}
\caption{FID scores and $SNR_{HO}$ computed by use of 3D ProGAN and 3D ProAmGAN-produced volumes.}
\label{table:metric}
\end{table}

\subsection{Style-AmbientGAN} 
\begin{figure}[H]
\centering
{\includegraphics[width=0.97\linewidth]{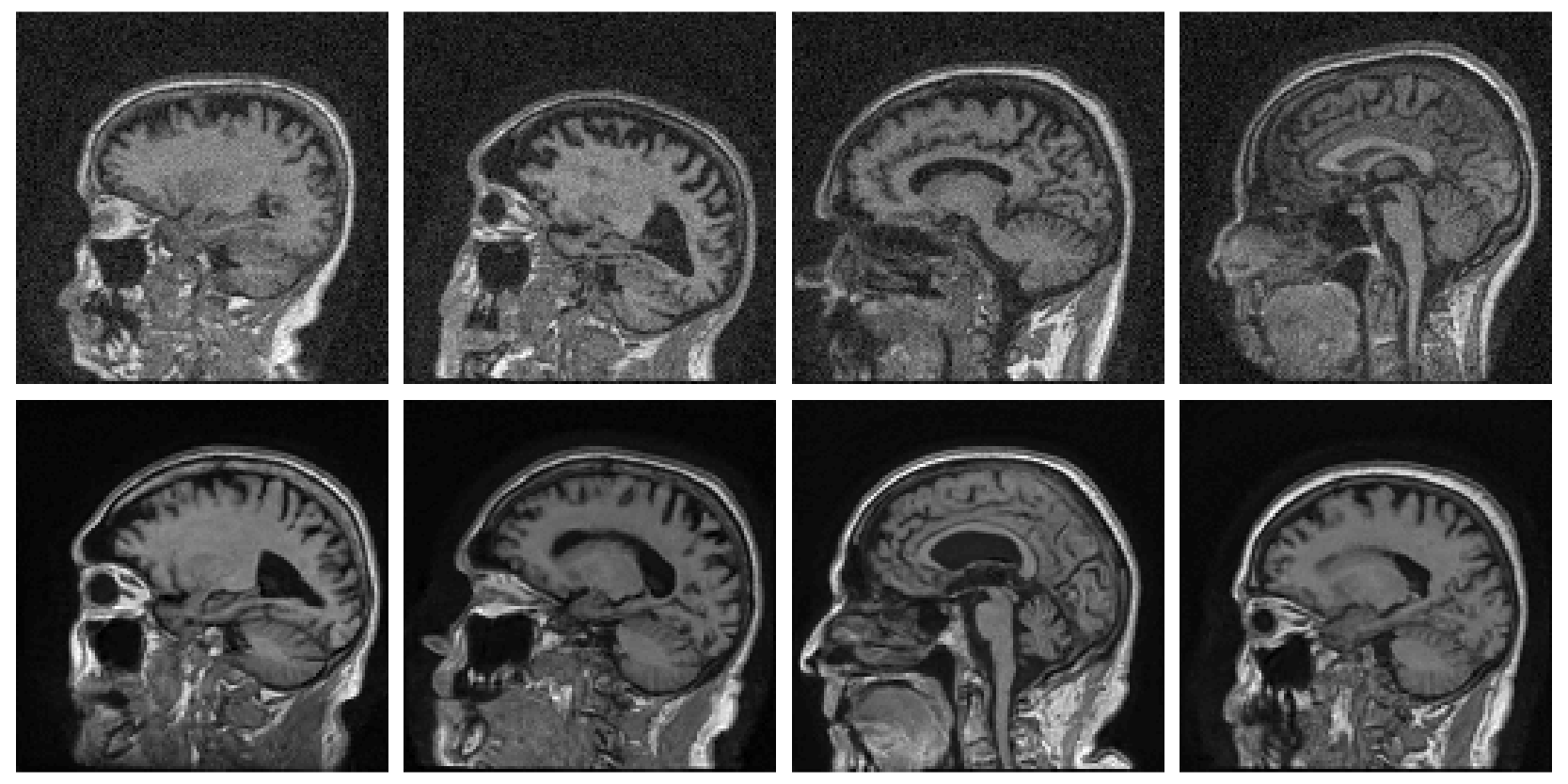}}
\caption{Top row: inverse DFT of the noisy measured data $\vec{g}$. Bottom row: StyAmGAN-generated images corresponding to different latent vectors.}
\label{fig:real_fakeStyle}
\end{figure}

\begin{figure}[H]
\centering
{\includegraphics[width=0.97\linewidth]{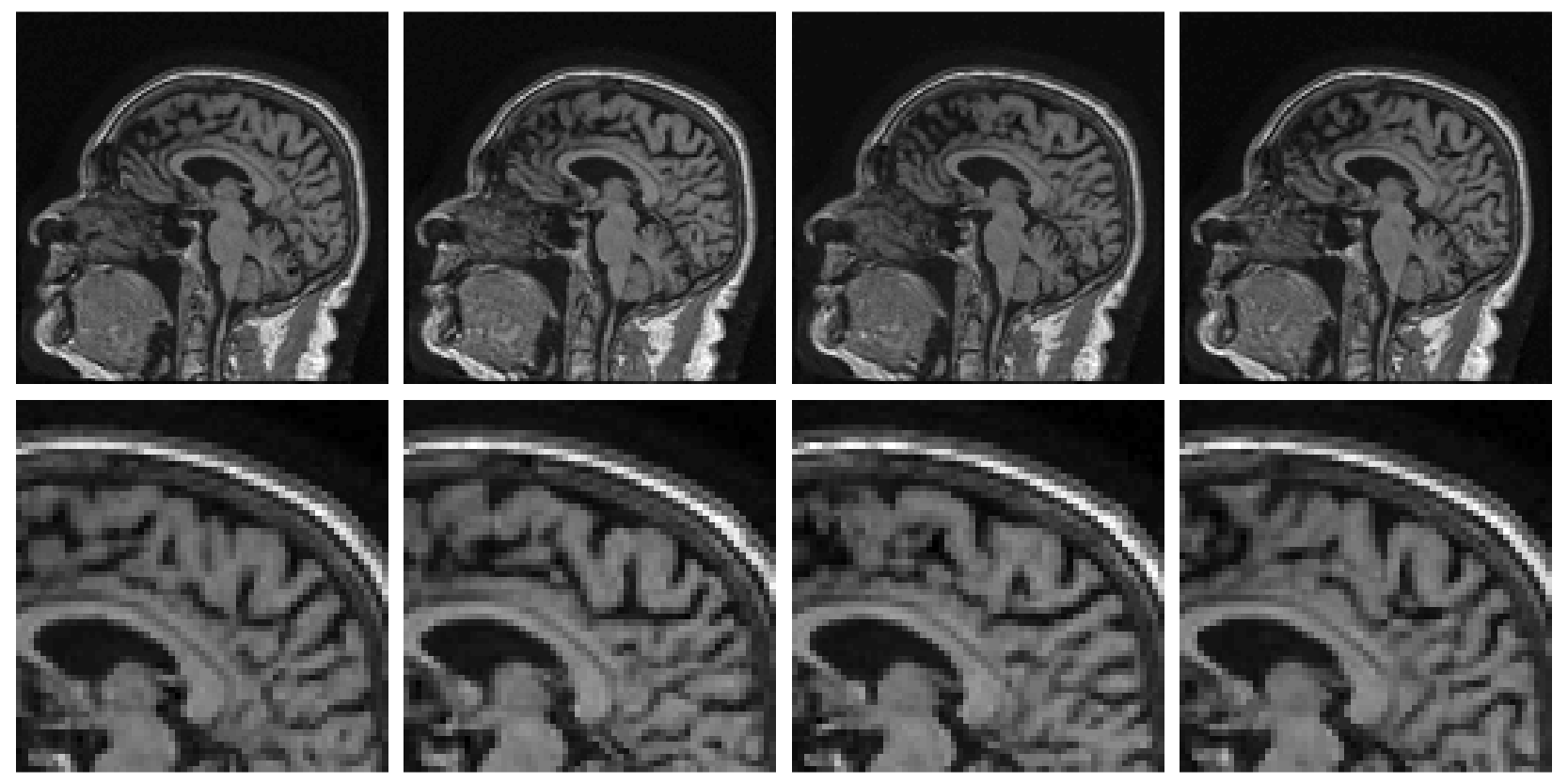}}
\caption{StyAmGAN-generated images with the same latent vector but different noise maps. Images in the bottom row are zoom-in regions of images in the top row.}
\label{fig:style}
\end{figure}

Examples of the noisy reconstructed images corresponding to the training measured data are shown in the top row of Fig. \ref{fig:real_fakeStyle}
and examples of StyAmGAN-generated images are shown in the bottom row of Fig. \ref{fig:real_fakeStyle}.
The StyAmGAN that is learned from noisy images can produce clean images,
which demonstrates the ability of StyAmGAN to mitigate the measurement noise when establishing SOMs. 
We investigated the ability of the StyAmGAN
to control scale-specific features in the synthesized objects. 
The use of style-based generator to manipulate fine-scale image textures while maintain the coarse-scale image structures was demonstrated.
Specifically, to generate images having the same large-scale structure but different fine-scale textures, the same latent vector and different noise maps were input to the style-based generator.
More details of the latent vector and the noise maps that form the input to the style-based generator can be found in the literature\cite{karras2019style, karras2019analyzing}.
Four images generated by the StyAmGAN with the same latent vector but different noise maps
are shown in Fig. \ref{fig:style}.

\section{Conclusion}
This study provides two important advances in training AmbientGANs for learning SOMs from imaging measurements: a 3D ProAmGAN for establishing 3D SOMs and a Style-AmbientGAN for controlling styles and features in the synthesized objects. Numerical studies considering stylized MR imaging systems were conducted. It was demonstrated that the 3D ProAmGAN can be successfully trained to produce high-quality 3D objects of $128\times128\times128$ voxels and the Style-AmbientGAN can provide ability to control specific styles and features in the synthesized objects. 
\vspace{-0.1cm}
\section*{ACKNOWLEDGMENT}       
This research was supported in part by NIH awards EB020604, EB023045, NS102213, EB028652, and NSF award DMS1614305.
\vspace{-0.1cm}
\bibliography{PAmbientGAN.bib}

\begin{thebibliography}{10}

\bibitem{barrett2013foundations}
Barrett, H.~H. and Myers, K.~J.,  [{\em Foundations of {I}mage
  {S}cience}{\nolinebreak\hspace{0.1em}]}, John Wiley \&amp; Sons (2013).

\bibitem{zhou2019approximating}
Zhou, W., Li, H., and Anastasio, M.~A., ``Approximating the {I}deal {O}bserver
  and {H}otelling {O}bserver for binary signal detection tasks by use of
  supervised learning methods,'' {\em IEEE {T}ransactions on {M}edical
  {I}maging}~{\bf 38}(10),  2456--2468 (2019).

\bibitem{zhou2020approximating}
Zhou, W., Li, H., and Anastasio, M.~A., ``Approximating the {I}deal {O}bserver
  for joint signal detection and localization tasks by use of supervised
  learning methods,'' {\em IEEE {T}ransactions on {M}edical {I}maging}~{\bf
  39}(12),  3992--4000 (2020).

\bibitem{zhou2020markov}
{Z}hou, W. and Anastasio, M.~A., ``{Markov-Chain Monte Carlo} approximation of
  the {I}deal {O}bserver using generative adversarial networks,'' in [{\em
  Medical Imaging 2020: Image Perception, Observer Performance, and Technology
  Assessment}{\nolinebreak\hspace{0.1em}]},   {\bf 11316},  113160D,
  International Society for Optics and Photonics (2020).

\bibitem{kupinski2003experimental}
Kupinski, M.~A., Clarkson, E., Hoppin, J.~W., Chen, L., and Barrett, H.~H.,
  ``Experimental determination of object statistics from noisy images,'' {\em
  JOSA A}~{\bf 20}(3),  421--429 (2003).

\bibitem{bora2018ambientgan}
Bora, A., Price, E., and Dimakis, A.~G., ``Ambientgan: {G}enerative models from
  lossy measurements,'' in [{\em International Conference on Learning
  Representations (ICLR)}{\nolinebreak\hspace{0.1em}]},  (2018).

\bibitem{zhou2019learning}
Zhou, W., Bhadra, S., Brooks, F., and Anastasio, M.~A., ``Learning stochastic
  object model from noisy imaging measurements using {AmbientGANs},'' in [{\em
  Medical Imaging 2019: Image Perception, Observer Performance, and Technology
  Assessment}{\nolinebreak\hspace{0.1em}]},   {\bf 10952},  109520M,
  International Society for Optics and Photonics (2019).

\bibitem{karras2017progressive}
Karras, T., Aila, T., Laine, S., and Lehtinen, J., ``Progressive {G}rowing of
  {GAN}s for improved quality, stability, and variation,'' {\em arXiv preprint
  arXiv:1710.10196}  (2017).

\bibitem{zhou2020progressively}
Zhou, W., Bhadra, S., Brooks, F.~J., Li, H., and Anastasio, M.~A.,
  ``{Progressively-Growing AmbientGANs} for learning stochastic object models
  from imaging measurements,'' in [{\em Medical Imaging 2020: Image Perception,
  Observer Performance, and Technology
  Assessment}{\nolinebreak\hspace{0.1em}]},   {\bf 11316},  113160Q,
  International Society for Optics and Photonics (2020).

\bibitem{zhou2020learning}
Zhou, W., Bhadra, S., Brooks, F.~J., Li, H., and Anastasio, M.~A., ``Learning
  stochastic object models from medical imaging measurements using
  {Progressively-Growing AmbientGANs},'' {\em arXiv preprint arXiv:2006.00033}
  (2020).

\bibitem{bhadra2020medical}
Bhadra, S., Zhou, W., and Anastasio, M.~A., ``Medical image reconstruction with
  image-adaptive priors learned by use of generative adversarial networks,'' in
  [{\em Medical Imaging 2020: Physics of Medical
  Imaging}{\nolinebreak\hspace{0.1em}]},   {\bf 11312},  113120V, International
  Society for Optics and Photonics (2020).

\bibitem{kelkar2020compressible}
Kelkar, V.~A., Bhadra, S., and Anastasio, M.~A., ``Compressible latent-space
  invertible networks for generative model-constrained image reconstruction,''
  {\em arXiv preprint arXiv:2007.02462}  (2020).

\bibitem{eklund2019feeding}
Eklund, A., ``Feeding the zombies: {S}ynthesizing brain volumes using a 3{D}
  progressive growing {GAN},'' {\em arXiv preprint arXiv:1912.05357}  (2019).

\bibitem{karras2019style}
Karras, T., Laine, S., and Aila, T., ``A style-based generator architecture for
  generative adversarial networks,'' in [{\em Proceedings of the IEEE
  Conference on Computer Vision and Pattern
  Recognition}{\nolinebreak\hspace{0.1em}]},   4401--4410 (2019).

\bibitem{karras2019analyzing}
Karras, T., Laine, S., Aittala, M., Hellsten, J., Lehtinen, J., and Aila, T.,
  ``Analyzing and improving the image quality of stylegan,'' {\em arXiv
  preprint arXiv:1912.04958}  (2019).

\end{thebibliography}
\bibliographystyle{spiebib} 

\end{document}